\documentclass[12pt]{article}
\usepackage{amsfonts,amssymb,amscd,amstext}
\usepackage{a4wide,theorem}

\makeatletter
\@addtoreset{equation}{section}
\makeatother
\renewcommand{\theequation}{\thesection.\arabic{equation}}
\newcommand\ie{\emph{i.e.}}
\newcommand\eg{\emph{e.g.}}
\newcommand\g{\ensuremath{\mathcal{G}}}

{
\theoremstyle{break}
\theorembodyfont{\normalfont}
\newtheorem{theorem}{Theorem}[section]

\newtheorem{proposition}{Proposition}[section]
\newtheorem{example}{Example}[section]
}

\newenvironment{proof}{\list{}{\setlength{\leftmargin}{0pt}}\item\relax
\noindent\emph{Proof.~}}{\endlist}

\newenvironment{noteadded}{\list{}{\setlength{\leftmargin}{0pt}}\item\relax
\noindent\emph{Note~added.~}}{\endlist}

\begin{document}

\thispagestyle{empty}
\vspace*{-1cm}
\begin{flushright}
FTUV--98/6
\quad
IFIC--98/6
\\
23 February, 1988
\\
(REVISED VERSION, 27 July, 1998)
\\
hep-th/9802192
\\[1.5cm]
\end{flushright}

\begin{center}
\begin{Large}
\bfseries{On the general structure of gauged Wess--Zumino--Witten terms}
\end{Large}

\vspace*{0.4cm}

\begin{large}
J. A. de Azc\'arraga
and
J. C. P\'erez Bueno\footnote{
E-mails: azcarrag@lie1.ific.uv.es, pbueno@lie.ific.uv.es}
\end{large}

\vspace*{0.4cm}
\begin{it}
Departamento de F\'{\i}sica Te\'orica, Universidad de Valencia
\\
and IFIC, Centro Mixto Universidad de Valencia--CSIC
\\
E--46100 Burjassot (Valencia), Spain
\end{it}

\end{center}

\begin{abstract}
The problem of gauging a closed form is considered.
When the target manifold is a simple Lie group $G$, it is seen that there is
no obstruction to the gauging of a subgroup $H\subset G$ if we may construct
from the form a
cocycle for the relative Lie algebra cohomology (or for the equivariant
cohomology), and an explicit general expression for these cocycles
is given. The common geometrical structure of the gauged 
closed forms and the D'Hoker and Weinberg effective actions of 
WZW type, as well as the obstructions for their existence, is also 
exhibited and explained.
\end{abstract}

\section{Introduction}
\label{sec.1}
Wess--Zumino--Witten (WZW) terms \cite{Wes.Zum:71,Wit:84,Wit:83} may be
described by closed forms on an
$n$--dimensional manifold $D$, the boundary $\partial D$ of which
is spacetime $M$.
More generally (as will be the case throughout this paper) they are given
by closed $n$--forms $\Omega$
on a certain manifold $P$, the target manifold, and it is
their pull--back to
$D$ by $\phi:D\to P$ which defines the integrand of the WZW
action, $I_{\text{WZW}} =\int_D \phi^*\Omega$.
Quite often, the target manifold is a Lie group $G$
\footnote{
It is also possible to define WZ terms $h$ on supergroups as, \emph{e.g.} on
superspace.
The Grassmann sector is topologically trivial, $h=db$ ($h$ is exact), and the
quasi--invariance of $b$ ($L_X h=0\Rightarrow \exists\gamma\ /\ L_X b =
d\gamma$)
becomes their characteristic property
(see \cite{Azc.Tow:89,Azc.Gau.Izq.Tow:89}).}.
WZW terms may be called topological in the sense that they depend on the
properties of the manifold on which they are defined (and not \emph{e.g.} on
the metric).
Since the variation of a closed form is in turn
closed, the classical Euler--Lagrange equations on the ($n-1$)--dimensional
spacetime are unambiguous when certain topological conditions (which will not
be of our concern here) are met (the quantum theory requires
\cite{Wit:84,Wit:83,DHo.Wei:94} the quantisation of the coefficient with
which $\Omega$ appears in the action).

The gauging of WZW terms, \ie\ the introduction of the
Yang--Mills fields via minimal coupling, does
not preserve the closedness condition and, as a result, it
is not always possible. The case of the two-dimensional sigma model
was solved in \cite{Jac.Jon.Moh.Osb:90}, and the possible obstructions
to the process of gauging rigid symmetries were discussed in
\cite{Wu:85,Hul.Spe:89,Hul.Spe:91,Pap:90,Wit:92b,Wu:93} and in
\cite{Fig.Sta:94},
which also included in its analysis the new topological terms in
\cite{deW.Hul.Roc:87}.
It was realised that the obstructions to the gauging process found in
\cite{Hul.Spe:91}
have an elegant geometric interpretation \cite{Wu:93,Fig.Sta:94}
in terms of equivariant cohomology
\cite{Ber.Ver:83,Ber.Get.Ver:91,Duf.Ver:93,Ati.Bot:84,Mat.Qui:86},
a fact that had also
been noticed by Witten \cite{Wit:92b}.
Equivariant cohomology has also appeared
\cite{Ouv.Sto.Baa:89,Kan:89,Nie.Tir:94} in other physical theories
of recent interest, and particularly in topological/cohomological
field theories \cite{Wit:91,Ati.Jef:90}
(for reviews and references see, \emph{e.g.},
\cite{Bir.Bla.Rak.Tho:91,Sto.Thu.Wal:93}).

Recently, D'Hoker and Weinberg \cite{DHo.Wei:94,DHo:95} have studied the
structure of the most general effective actions
with symmetry group $G$ broken down to a subgroup $H$.
These include those based on $G$--invariant Lagrangian densities
\cite{Wei:68,Col.Wes.Zum:69,Cal.Col.Wes.Zum:69,Sal.Str:69} as well as those
given by
quasi--invariant Lagrangians (which change by a total derivative) as is the
case of the WZW terms.
It has been pointed
out \cite{Azc.Mac.Per:98} how, for $G$ simple, it is possible to derive a
general
expression for the effective actions of WZW type by looking for cocycles of
the relative Lie algebra cohomology 
$H^*(\mathcal{G},\mathcal{H};\mathbb{R})$.

We show explicitly in this paper that the existence of both types
of action terms, gauged WZW terms and the effective actions of WZW type
of D'Hoker and Weinberg, have the same cohomological origin.
As a result, the obstructions encountered in their construction
are given in terms of the same type of `anomaly'. This common
structure is the result of the mathematical equivalence of the
classes of projectable closed invariant forms, the relative Lie 
algebra cohomology $H^*(\mathcal{G},\mathcal{H};\mathbb{R})$
(see \cite{Che.Eil:48,Est:55})
and the equivariant cohomology $H^{*}_H (G)$. Thus,
this paper may also be seen as a proof of these mathematical 
equivalences in terms of physical theories.
To make this explicit, we shall first recover 
in Secs.~\ref{sec.2} and \ref{sec.3} the obstructions to
the gauging process \cite{Hul.Spe:91} in terms of equivariant cohomology,
but using the Kalkman BSRT operator \cite{Kal:93} rather than the
Mathai-Quillen
\cite{Mat.Qui:86} isomorphism used in \cite{Wu:93,Fig.Sta:94}.
We shall then find in Sec.~\ref{sec.4}
the general form of an $H$--gauged WZW term with target manifold $G$,
and discuss in Sec.~\ref{sec.5} why this problem and that of 
finding effective actions on a coset manifold are similar,
making the comparison explicit.
Sec.~\ref{sec.6} is devoted to exploit the larger (left$\times$right)
$G\times G$ symmetry of the
cocycles on a compact group $G$, and to give a general expression of the
gauged WZW term which is then applied to an example.
Some calculations are indicated in the Appendix;
the mathematical background for the paper is summarised in the 
next section.

\section{Mathematical preliminaries. The Weil algebra}
\label{sec.1a}

Let $P(H,K)$ be a principal bundle with structure group $H$ acting on $P$ from
the right and let
$\{X_\alpha\}$ ($\{\omega^\alpha\},\
\omega^\alpha(X_\beta)=\delta_\beta^\alpha$)
($\alpha,\beta=1,\dots,\text{dim}\,{H}$) be a basis for its Lie algebra
$\mathcal{H}$ (dual $\mathcal{H}^*$).
Let $\mathcal{W}(\mathcal{H})$ be
the Weil algebra \cite{Car:50,Wei:49,Ati.Bot:84,Mat.Qui:86},
$\mathcal{W}(\mathcal{H})=
\wedge(\mathcal{H}^*)\otimes\mathcal{S}(\mathcal{H}^*)$,
where $\wedge(\mathcal{H}^*)$
is the algebra of multilinear antisymmetric mappings on $\mathcal{H}$ and
$\mathcal{S}(\mathcal{H}^*)$ the symmetric algebra on
$\mathcal{H}^*$ (or symmetric polynomials on $\mathcal{H}$).
Endowed with the Weil differential $d_W$, $\mathcal{W}(\mathcal{H})$ becomes a
differential graded commutative algebra freely generated by the elements
$\theta^\alpha$ (of degree 1) and $u^\alpha$ (of degree 2) in
$\wedge(\mathcal{H}^*)$ and
$\mathcal{S}(\mathcal{H}^*)$ respectively, satisfying the relations
\begin{equation}
d_W\theta^\alpha +
{1\over 2} C^{\alpha}_{\beta\gamma}\theta^\beta\wedge \theta^\gamma = u^\alpha
\ ,\ {}
d_W u^\alpha = - C^{\alpha}_{\beta\gamma} \theta^\beta \wedge u^\gamma
\ ;\ {}
i_{W\,X_\alpha}\theta^\beta=\delta_\alpha^\beta
\ ,\ {}
i_{W\,X_\alpha} u^\beta=0\ ;
\label{1.1a}
\end{equation}
$d^2_W=0$ follows from the Jacobi identity.
An element in $\wedge^q(\mathcal{H}^*)\otimes\mathcal{S}^s(\mathcal{H}^*)$
has degree $(q+2s)$.

Let $A$ be a connection on $P$ and $F$ 
its curvature; $A$ and $F$ are $\mathcal{H}$--valued forms
$A=A^\alpha X_\alpha$, $F=F^\alpha X_\alpha$. Eqs. (\ref{1.1a})
are the same as those satisfied by $A^\alpha$ and $F^\alpha$.
Thus, the mapping
$\phi_W:(\theta^\alpha,u^\alpha) \mapsto (A^\alpha,F^\alpha)$
induces a homomorphism
$\phi_W: \mathcal{W}(\mathcal{H}) \to \wedge(P)$
of differential algebras which is called the \emph{Weil homomorphism}
determined by the connection $A$ on $P$.
As a result, the Weil algebra provides a 
universal model for the relations satisfied
by any connection $A$ and curvature $F$ on $P(H,K)$,
\begin{equation}
\begin{array}{c}
\displaystyle
d_W A ^\alpha + {1\over 2}
C^{\alpha}_{\beta\gamma} A^\beta \wedge A ^\gamma
= F ^\alpha \quad
(F = d_WA + {1\over 2}[A,A])\quad,
\\[0.3cm]
\displaystyle
d_W F ^\alpha = - C^{\alpha}_{\beta\gamma} A^\beta \wedge F^\gamma
\quad
(d_WF + [A,F]=0)\quad,
\\[0.3cm]
\displaystyle
i_{W\,X_{\alpha}} A^\beta = \delta_\alpha^\beta
\quad,\quad
i_{W\,X_{\alpha}} F^\beta = 0\quad,
\end{array}
\label{1.1}
\end{equation}
where we keep $i_W$ and $d_W$ to denote the corresponding differential
operators acting on $A$, $F$, cf. (\ref{1.1a}).
Forms which are $H$--invariant and horizontal ($i_{X_\alpha}(\ )=0$) are called
\emph{basic}.
Thus, the forms on the base manifold $K$ may be identified with the 
basic forms on the total manifold $P$, {\it i.e.} with those
that are `{\it projectable}' to $K$ (a term which, more 
precisely, indicates that the bundle projection $\pi$ induces an 
embedding $\pi^*:\wedge(K)\to\wedge(P)$
which determines the basic forms on $P$).

The Weil homomorphism is compatible with the differentials, the contraction and
the action of $H$ ($[\phi_W,d_W]=0,\ [\phi_W,i_W]=0,\ 
[\phi_W,L_{W\,X}]=0$ where $L_{W\,X}$ is the Lie derivative
with respect the vector field $X$)
\cite{Car:50,Wei:49,Ati.Bot:84,Mat.Qui:86}.
Using (\ref{1.1}),
$L_{W\,X} = i_{W\,X} d_W + d_W i_{W\,X}$ gives
\begin{equation}
L_{W\,X_\alpha} A^\beta = - C^{\beta}_{\alpha\gamma} A ^\gamma
\ {}
(L_{W\,X} A = - [X , A])
\ ,\ {}
L_{W\,X_\alpha} F^\beta = - C^{\beta}_{\alpha\gamma} F^\gamma
\ {}
(L_{W\,X} F = - [X, F])
\quad;
\label{1.2}
\end{equation}
$[L_{W\,X_\alpha},i_{W\,X_\beta}] = i_{W\,[X_\alpha,X_\beta]}$. 
Since
\begin{equation}
\label{local}
L_{\zeta^\alpha X_\alpha} = \zeta^\alpha L_{X_\alpha} +
d\zeta^\alpha \wedge i_{X_\alpha}\quad,
\end{equation}
we see that
\begin{equation}
L_{W\,\zeta^\alpha X_\alpha} A^\beta = d\zeta^\beta -
\zeta^\alpha C^{\beta}_{\alpha\gamma} A ^\gamma
\quad,\quad
L_{W\,\zeta^\alpha X_\alpha} F^\beta =
- \zeta^\alpha C^{\beta}_{\alpha\gamma} F^\gamma
\label{1.3}
\end{equation}
\ie, the action of $L_{W\,\zeta^\alpha X_\alpha}$ 
on $A$ and $F$ generates the gauge
transformation $\delta_{\zeta}$ associated with the group parameters
$\zeta^\alpha$ of $H$.
If we add the zero and one forms $\zeta^\alpha$ and $d\zeta^\alpha$ to the
generators $A,\ F$ of the Weil algebra, the resulting one (as
$\mathcal{W}(\mathcal{H})$ itself)
is a contractible \cite{Sul:77} free differential algebra,
and hence it has trivial de Rham cohomology (see also \cite{Car:50}).
If we have (matter) fields $\varphi^i$ defined 
on $P$ or through some associated bundle on which $H$ acts,
$ \delta_\zeta\varphi^i = L_{\zeta^\alpha X_\alpha} \varphi^i =
\zeta^\alpha L_{X_\alpha} \varphi^i$.
For a linear action $T_{\alpha}$, we have
\begin{equation}
L_{\zeta^\alpha X_\alpha} \varphi^i =
- \zeta^\alpha (T_\alpha)_{\cdot j}^i \varphi^j
\quad
\text{with}\
X_\alpha= X_\alpha^i(\varphi){\partial\over \partial \varphi^i}\ ,\
X_\alpha^i (\varphi) = (T_\alpha)_{\cdot j}^i \varphi^j\quad,
\label{1.4}
\end{equation}
where $T$ is in the representation of ${\mathcal H}$ provided
by the fields $\varphi^i$.
Strictly speaking, the gauge transformations are 
not (\ref{1.3}), (\ref{1.4}),
but their pull backs to a suitable spacetime.
Nevertheless we can use these expressions to discuss the universal
obstructions to the gauging process (\ie, there 
will have a solution if these obstructions are absent).

The Lie derivative property (\ref{local}) shows 
immediately why horizontality 
plays an essential role in the discussion:
if the form is horizontal, the term containing $i_{X_\alpha}$ will not
contribute, and $L_{\zeta^\alpha X_\alpha}$ will be given by
$\zeta^\alpha L_{X_\alpha}$ even if the parameters $\zeta^\alpha$
are not constant. For instance,
$L_{\zeta^\alpha X_\alpha} \varphi^i= \zeta^\alpha
L_{X_\alpha} \varphi^i$ but this is of course not the case for
$d\varphi^i$ (see (\ref{dphi}) below).

\emph{A comment on notation}.
Being $A$ a connection on the principal bundle $P(H,K)$,
$A$ and $F$ are forms on the manifold $P$, and 
hence in $\wedge(P)$.
However, for the purposes of this paper and to make the gauge mechanism
clearer, it is practical to consider $A$ and $F$ as the generators of a
separate algebra.
Since $A,\, F$ are a copy of the generators of the universal Weil algebra,
we shall treat them as generators of $\mathcal{W}(\mathcal{H})$.
In this way, a form with components `in $\mathcal{W}(\mathcal{H})$'
will indicate
that it includes terms in the connection $A$ and/or curvature $F$,
and a form
`in $\wedge(P)$' will refer to an ungauged form, with no components in the
Yang--Mills fields or strengths
(alternatively, we could keep the generators $\theta^\alpha$ and $u^\alpha$
of $\mathcal{W}(\mathcal{H})$ in (\ref{1.1a})
throughout and replace them by $A$, $F$ using $\phi_W$ at the end).
In the above framework, it will be convenient to distinguish between the
operators on $\mathcal{W}(\mathcal{H})$ and those acting on
$\wedge(P)$.
To this aim, we shall keep the subindex $\scriptstyle W$ for the operators
in (\ref{1.1}), and reserve the notation $d,\ i_{X_\alpha}$ (or
$i_{\alpha}$) for
their counterparts acting on $\wedge(P)$ or on the exterior algebra on an
associated bundle.
The total $\mathbf{d}$
and $\mathbf{i}$ will denote the sums $\mathbf{d}=d_W\otimes 1 + 1\otimes d$
and $\mathbf{i}_\alpha = i_{W\alpha}\otimes 1 + 1\otimes i_\alpha$
($\mathbf{d}^2=0=\mathbf{i}^2$) acting on
$\mathcal{W}(\mathcal{H})\otimes\wedge(P)$.
Similarly,
$\mathbf{L}_\alpha \equiv \mathbf{i}_\alpha \mathbf{d} +
\mathbf{d} \mathbf{i}_\alpha = L_{W\alpha}\otimes 1 + 1\otimes L_\alpha$;
on \emph{horizontal} forms
$\mathbf{L}_{\zeta^\alpha X_\alpha} = \zeta^\alpha \mathbf{L}_{\alpha}$.
The $\otimes$ symbol will be often omitted if no confusion arises.

\section{Gauging closed forms}
\label{sec.2}

Let $\Omega$ be an $n$--form on $P$ (or on an associated bundle) and
let $\varphi^i$ be the coordinates of $P$.
Let $H$ be the compact and simply connected group to be gauged.
The {\it  minimal coupling} substitution
has the form
\begin{equation}
d\mapsto D:= d- A^\alpha L_\alpha\quad,
\label{2.1}
\end{equation}
(\ie, $1\otimes d\to 1\otimes d - A^\alpha \otimes L_\alpha$), where
$L_\alpha$ is the Lie derivative with respect to the vector field
associated with the right action of $H$ on $P$
($X_{\alpha} = X_\alpha^i(\varphi)\partial / \partial\varphi^i,
\ L_\alpha\varphi^i  = X_\alpha^i(\varphi)$).
Indeed, we may check that in the present language

\begin{equation}
\begin{array}{@{}r@{}l}
\mathbf{L}_{\zeta^\alpha X_\alpha} D\varphi
&
=
(d\zeta^\alpha \wedge \mathbf{i}_\alpha + \zeta^\alpha \mathbf{L}_\alpha)
(d -A^\beta L_\beta)\varphi
\\[0.3cm]
& =
(d\zeta^\alpha \wedge i_\alpha d
-d\zeta^\alpha \wedge (i_{W\,\alpha} A^\beta) L_\beta
+d\zeta^\alpha \wedge A^\beta i_\alpha L_\beta )\varphi
+\zeta^\alpha \mathbf{L}_\alpha (D\varphi)
= \zeta^\alpha \mathbf{L}_\alpha (D\varphi) \ .
\end{array}
\label{dphi}
\end{equation}
Under (\ref{2.1}),
$\Omega\mapsto \widetilde\Omega$ \ie,
\begin{equation}
\Omega ={1\over n!} \Omega_{i_1 \dots i_n} d\varphi^{i_1}\wedge\dots\wedge
d\varphi^{i_n}
\mapsto
\widetilde \Omega ={1\over n!} \Omega_{i_1 \dots i_n}
D\varphi^{i_1} \wedge\dots\wedge D\varphi^{i_n}
\quad.
\label{2.2}
\end{equation}

To see how the minimal coupling affects the closedness of $\Omega$
we have to compute $\mathbf{d}\widetilde\Omega$. Using
\begin{equation}
\mathbf{d}D\varphi^i=
-F^\alpha L_\alpha \varphi^i + A^\alpha D L_\alpha \varphi^i =
-F^\alpha i_\alpha d \varphi^i + A^\alpha D L_\alpha \varphi^i \quad,
\label{2.4}
\end{equation}
we find
\begin{equation}
\begin{array}{l}
\displaystyle
\mathbf{d}\widetilde\Omega = \widetilde{d\Omega} + {1\over n!} \Bigl\{
\partial_j \Omega_{i_1 \dots i_n} A^\alpha L_\alpha \varphi^j \wedge
D \varphi^{i_1} \wedge \dots \wedge D \varphi^{i_n} \\
\displaystyle
\qquad
+ \sum_{s=1}^n (-1)^{s+1} \Omega_{i_1 \dots i_n} [A^\alpha D L_\alpha
\varphi^{i_s} - F^\alpha i_\alpha d \varphi^{i_s}]  D \varphi^{i_1} \wedge
\dots \wedge \widehat{D \varphi^{i_s}} \wedge \dots \wedge D \varphi^{i_n}
\Bigr\}
\end{array}
\label{2.5}
\end{equation}
where, in general, the tilde indicates the result of performing the minimal
substitution (\ref{2.1}) in the expression underneath.
The last term in (\ref{2.5}) is clearly $-F^\alpha \widetilde{i_\alpha \Omega}$
($i_\alpha d\varphi=\widetilde{i_\alpha d\varphi}$)
and the second and the third are easily identified with
$A^\alpha \widetilde{L_\alpha \Omega}$. Hence,
\begin{equation}
\mathbf{d} \widetilde\Omega =
\widetilde{d\Omega} + A^\alpha \widetilde{L_\alpha \Omega}
- F^\alpha \widetilde{i_\alpha \Omega} =
\widetilde{ d\Omega} +
\widetilde{ A^\alpha L_\alpha \Omega} - \widetilde{F^\alpha i_\alpha \Omega}
\label{2.6}
\end{equation}
which is \cite[eq. (4.2)]{Hul.Spe:91}.
Since the different terms in (\ref{2.6}) are independent, it follows that a
closed form $\Omega$ will remain closed after gauging the group $H$ iff
\begin{description}
\item[a)] it is horizontal ($i_\alpha\Omega = 0$)
\item[b)] $\Omega$ is invariant under the right translations of $H$ generated
by the vector fields $X_{\alpha}\in\mathcal{H}$.
\end{description}
If $\Omega$ satisfies a) \emph{and} b), $d\Omega$ also satisfies them.
These are also the conditions that guarantee the existence of WZW--type
effective
actions on coset spaces \cite{DHo.Wei:94,DHo:95,Azc.Mac.Per:98} and will
explain the formal similarity of their general expressions in
\cite{Azc.Mac.Per:98} with those which will be found later for the present
case.

However if a form $\Omega$ is $(1\otimes i_\alpha)$--horizontal the minimal
coupling (\ref{2.1})
does not act since $D\varphi^i=d\varphi^i-A^\alpha L_\alpha
d\varphi^i=(1-A^\alpha i_\alpha)d\varphi^i$ and $\widetilde\Omega=\Omega$.
In fact, a horizontal and $H$--invariant form is automatically gauge invariant.
Thus, to obtain a non--trivial result and incorporate the Yang--Mills fields
we need `extending' $\Omega$ to a form
$\widetilde\beta\in \mathcal{W}(\mathcal{H})\otimes \wedge(P)$ such that
$\widetilde\beta(A=0,F=0)=\Omega$.
In this case, (\ref{2.6}) is trivially modified to read
\begin{equation}
\mathbf{d}\widetilde\beta =
\widetilde{\mathbf{d}\beta} +
\widetilde{ A^\alpha L_\alpha \beta} - \widetilde{F^\alpha i_\alpha \beta}
\equiv \widetilde{\delta\beta}
\quad,
\label{2.6a}
\end{equation}
where
\begin{equation}
\delta:= d_W \otimes 1 + 1\otimes d + A^\alpha \otimes L_\alpha -
F^\alpha \otimes i_\alpha
\quad
(\delta := \mathbf{d} + A^\alpha L_\alpha - F^\alpha i_\alpha)\quad;
\label{2.7}
\end{equation}
it may be easily checked that $\delta^2=0$.
This is the BRST operator of \cite{Kal:93} which we now discuss in the present
context.

\section{The Mathai--Quillen and Kalkman isomorphisms\\
and the gauging of forms}
\label{sec.3}

The minimal coupling (\ref{2.1}) defines a one--to--one
correspondence,
the \emph{gauging map} $\psi:\beta\mapsto\widetilde\beta$,
between ungauged ($\beta$) and gauged ($\widetilde\beta$) forms.
Since $\mathbf{d}\widetilde\beta =
\psi (\delta\beta)$ (eq. (\ref{2.6a})),
$\psi^{-1} \mathbf{d}\psi \beta =\delta\beta$.
Hence
\begin{equation}
\psi^{-1} \mathbf{d} \psi = \delta = d_W \otimes 1 + 1\otimes d
+ A^\alpha \otimes L_\alpha - F^\alpha \otimes i_\alpha
\quad.
\label{3.1}
\end{equation}
The map $\psi$ is given by \cite{Kal:93} (cf. \cite{Mat.Qui:86})
\begin{equation}
\psi=\exp\left(-\sum_{\alpha=1}^{\text{dim}\,\mathcal{H}}
A^\alpha\otimes i_\alpha\right)
= \prod _\alpha (1 - A^\alpha\otimes i_\alpha)\quad,
\label{3.2}
\end{equation}
where in the last term there is no sum in $\alpha$;
$\displaystyle\psi^{-1}=\exp\Big(\sum_\alpha A^\alpha\otimes i_\alpha\Big)$.
Eq. (\ref{3.1}) may be checked using
$\psi^{-1} \mathbf{d} \psi = \exp (ad(A^\alpha\otimes i_\alpha)) \mathbf{d}$
and the relations
\begin{equation}
\begin{array}{l}
[A^\alpha \otimes i_\alpha, \mathbf{d}] =
- d A^\alpha \otimes i_\alpha + A^\alpha
\otimes L_\alpha \quad,\quad
[A^\alpha \otimes i_\alpha, d A^\alpha \otimes i_\alpha] = 0 \quad,
\\[0.25cm]{}
[A^\alpha \otimes i_\alpha, A^\beta \otimes L_\beta] = -
C_{\beta\alpha}^\gamma A^\beta A^\alpha \otimes i_\gamma \quad,
\\[0.25cm]{}
[A^\alpha \otimes i_\alpha, [A^\beta \otimes i_\beta, \mathbf{d}]] = -
C_{\beta\alpha}^\gamma A^\beta A^\alpha \otimes i_\gamma \quad
\end{array}
\end{equation}
(higher order terms are zero), and taking into account that, in
$\mathcal{W}(\mathcal{H})\otimes \wedge(P)$,
$(u_1\otimes v_1)(u_2\otimes v_2)=(-1)^{v_1 u_2} (u_1 u_2 \otimes v_1 v_2)$.
As an example of the action of $\psi$ we may check easily that, on $d\varphi$
[\ie, on $(1\otimes d)(1\otimes\varphi)$],
$\psi: d\varphi\mapsto D\varphi$ since
$\displaystyle \prod_\alpha (1-A^\alpha \otimes i_\alpha) d\varphi$
(no sum in $\alpha$) is given (restoring the summation convention) by
$d\varphi - A^\alpha \otimes i_\alpha d\varphi =
(d - A^\alpha \otimes L_\alpha) \varphi$
\ie, by
$(d- A^\alpha L_\alpha)\varphi = D\varphi$.
Hence (\ref{3.2}) implements the minimal coupling.

Let us now take two copies $\mathfrak{A},\ \mathfrak{B}$ of the algebra
$\mathcal{W}(\mathcal{H})\otimes\wedge(P)$ endowed with the
differential operators $\delta$ and $\mathbf{d}$ respectively.
($\mathfrak{A},\delta$)
and ($\mathfrak{B},\mathbf{d}$) are not equal as \emph{differential} algebras,
but $\psi:\mathfrak{A}\to \mathfrak{B}$ makes them
isomorphic \cite{Kal:93}. Thus, their cohomology rings coincide,
$H^*_\delta(\mathfrak{A})=H^*_{\mathbf{d}}(\mathfrak{B})$: if $\beta \in
\mathfrak{A}$ and
$\delta \beta =0$, then $\mathbf{d}\widetilde\beta=0$ for
$\psi\beta =\widetilde{\beta}\in \mathfrak{B}$.
Moreover, these rings are both equal to $H_{\text{DR}}(P)$ because, being
contractible, the
$\mathcal{W}(\mathcal{H})$ part in ($\mathfrak{B},\mathbf{d}$) has trivial de
Rham cohomology.

Let us go back to (\ref{3.1}) and eqs. (\ref{1.3}), (\ref{1.4})
and restrict $\mathfrak{B}$ to the subalgebra of the
horizontal and invariant (hence gauge invariant) forms.
These forms $\widetilde\alpha$ fulfil the conditions
\begin{equation}
\mathbf{i}_\alpha \widetilde\alpha \equiv
( i_{W\alpha} \otimes 1 + 1\otimes i_\alpha)\widetilde\alpha=0
\quad,\quad
\mathbf{L}_\alpha \widetilde\alpha \equiv
(L_{W\alpha} \otimes 1 + 1\otimes L_\alpha)\widetilde\alpha =0 \quad,
\label{3.4}
\end{equation}
which are also satisfied by $\mathbf{d}\widetilde\alpha$,
since $\mathbf{L}_\alpha = \mathbf{d}  \mathbf{i}_\alpha +
\mathbf{i}_\alpha \mathbf{d}$.
Thus, these forms constitute a \emph{subalgebra} of
$(\mathcal{W}(\mathcal{H})\otimes \wedge(P),\mathbf{d})$,
the subalgebra of \emph{basic} forms
$([\mathcal{W}(\mathcal{H})\otimes \wedge(P)]_{\text{basic}},\mathbf{d})$ of
the
\emph{Weil model} for the equivariant cohomology $H_H^*(P)$.
It is easy to check that \footnote{These expressions follow 
by noticing that $[A^\alpha \otimes i_\alpha, 
i_{W\beta}\otimes 1 + 1\otimes i_\beta]=
-1\otimes i_\beta$,$\;$ $[A^\alpha \otimes i_\alpha,1\otimes i_\beta]=0$
and $[A^\alpha \otimes i_\alpha,  L_{W\beta} \otimes 1 
+ 1\otimes L_\beta]=0$.}

\begin{equation}
\psi^{-1} (i_{W\alpha}\otimes 1 + 1\otimes i_\alpha)\psi =
i_{W\alpha}\otimes 1
\quad,\quad
[\psi, L_{W\alpha} \otimes 1 + 1\otimes L_\alpha] =0
\quad.
\label{3.4a}
\end{equation}
Hence, the algebra
$([\mathcal{W}(\mathcal{H})\otimes \wedge(P)]_{\text{basic}},\mathbf{d})$ is
isomorphic to the $H$--invariant subalgebra
$([\mathcal{S}(\mathcal{H}^*)\otimes \wedge(P)]^H,d_C)$
of the \emph{Cartan model}, where
\begin{equation}
d_C= 1\otimes d - F^\alpha \otimes i_\alpha \quad ,
\label{dcartan}
\end{equation}
(or, simply, $d - F^\alpha i_\alpha$);
on $[\mathcal{S}(\mathcal{H}^*)\otimes \wedge(P)]^H$, $d_C^2=0$.
This is the Mathai--Quillen isomorphism \cite{Mat.Qui:86} and
$([\mathcal{S}(\mathcal{H}^*)\otimes \wedge(P)]^H,d_C)$
is the complex for the Cartan model of equivariant cohomology \cite{Car:50}.
The expression of $d_C$ follows from (\ref{3.1}) restricting it to horizontal
forms:
since $(i_{W\alpha}\otimes 1)\alpha=0$, $d_W = A^\alpha L_{W\alpha}$
($A^\alpha L_{W\,\alpha} F^\beta = - A^\alpha C_{\alpha \gamma}^\beta F ^\gamma
=d_{W} F^\beta$).
Then, since $\mathbf{L}_\alpha \widetilde\alpha=0$, $\delta$ reduces to
$d_C$.
The above results may be summarised in the diagram
\begin{equation}
\begin{array}{c}
\displaystyle
\text{[\emph{Intermediate scheme}]}
\\[0.3cm]
\displaystyle
\begin{CD}
& (\mathfrak{A}=\mathcal{W}(\mathcal{H})\otimes \wedge(P),\delta)
& \mathop{{\displaystyle \longrightarrow\atop \displaystyle \longleftarrow}}
\limits_{\psi^{-1}}^{\psi} &
(\mathfrak{B}=\mathcal{W}(\mathcal{H})\otimes \wedge(P),\mathbf{d})
\\
& @AAA
@AAA
\\
{(i_{W\alpha}\otimes 1)
\atop
(L_{W\alpha} \otimes 1 + 1\otimes L_\alpha)}\quad
& ([\mathcal{S}(\mathcal{H}^*)\otimes\wedge(P)]^H,d_C)
&\ \mathop{{\displaystyle \longrightarrow\atop \displaystyle \longleftarrow}}
\limits_{\psi^{-1}}^{\psi}\ &
([\mathcal{W}(\mathcal{H})\otimes \wedge(P)]_{\text{basic}},\mathbf{d})
&
\quad {(i_{W\alpha}\otimes 1 + 1\otimes i_\alpha)
\atop
(L_{W\alpha} \otimes 1 + 1\otimes L_\alpha)}
\\
& \text{[\emph{Cartan model}]} & & \text{[\emph{Weil model}]}
\end{CD}
\end{array}
\label{3.4b}
\end{equation}
It is easy to see that the image of
$\widetilde\alpha\in
[\mathcal{W}(\mathcal{H})\otimes \wedge(P)]_{\text{basic}}$
in $[\mathcal{S}(\mathcal{H}^*)\otimes\wedge(P)]^H$
by ${\psi^{-1}}$ is obtained by setting $A^\alpha=0$ \cite{Kal:93} in
$\widetilde\alpha$.
Let $\widetilde\alpha(A=0)\equiv\alpha$.
Then the previous analysis shows that the $\mathbf{d}$--closed elements which
determine the $n$--cocycles of $H_H^n(P)$ for the Weil model
are represented in the Cartan model by $d_C$--cocycles in
${\sum_{s\,\oplus}}
[\mathcal{S}^s(\mathcal{H}^*)\otimes\wedge_{n-2s}(P)]^H$.
A $d$--closed form $\Omega\in\wedge(P)$ will be gaugeable
\cite{Wit:92b,Wu:93,Fig.Sta:94} iff it admits
an equivariant extension
$\alpha\in [\mathcal{S}(\mathcal{H}^*)\otimes\wedge(P)]^H$
in the Cartan model ($d_C \alpha=0$). The gauged, closed and gauge invariant
form is then the associated $\mathbf{d}$--cocycle
$\widetilde{\alpha}=\psi(\alpha)$ in the Weil model.

Let $\Omega$ be a closed $n$--form.
As we have seen, performing in it
the minimal substitution (\ref{2.1}) does not solve the problem
of gauging $\Omega$ as it stands.
However, let $\alpha\in\mathcal{S}(\mathcal{H}^*)\otimes\wedge(P)$
be the $(i_{W\,\alpha}\otimes 1$--horizontal) form
\begin{equation}
\alpha = \Omega + \sum_{s=1}^p F^{\alpha_1}\wedge\dots\wedge F^{\alpha_s}
v_{\alpha_1\dots \alpha_s}
\equiv
\Omega + \sum_{s=1}^p v^{(s,n-2s)}
\quad,\quad
\alpha(F=0)=\Omega\quad,
\label{3.5}
\end{equation}
where $p$ is the integer part of $n/2$ and $v_{\alpha_1\dots \alpha_s}$ is a
$(n-2s)$--form on $P$,
\begin{equation}
v_{\alpha_1\dots \alpha_s}= {1\over (n-2s)!}
v_{\alpha_1\dots \alpha_s j_1\dots j_{n-2s}}
d\varphi^{j_1}\wedge\dots\wedge d\varphi^{j_{n-2s}}
\quad.
\label{3.6}
\end{equation}
If
$\alpha\in [\mathcal{S}(\mathcal{H}^*)\otimes\wedge(P)]^H$,
the $\wedge_{n-2s}(P)$--valued symmetric polynomials
$v_{\alpha_1 \dots \alpha_s}$ must be $H$--invariant,
$\mathbf{L}_\beta v^{(s,n-2s)}=0$, \ie,
\begin{equation}
\mathbf{L}_\beta v^{(s,n-2s)}=
(L_\beta v_{\alpha_1\dots \alpha_s}
- C_{\beta\alpha_1}^\gamma v_{\gamma \alpha_2\dots \alpha_s}
\dots
- C_{\beta\alpha_s}^\gamma v_{\alpha_1\dots \alpha_{s-1}\gamma})
F^{\alpha_1}\wedge\dots\wedge F^{\alpha_s} = 0
\quad.
\label{3.6a}
\end{equation}
Since the second part in (\ref{3.6a}) is simply
$(\text{coad} X_\beta)^{\otimes s}$,
we see that our $\mathbf{L}_\beta$ may be identified with the
`covariant derivative' in \cite{Hul.Spe:91},
$L_\beta^{\text{cov}}:= L_\beta + (\text{coad} X_\beta)^{\otimes s}$.
The cocycle condition
$d_C\alpha =(d-F^\beta i_\beta)\alpha=0$ now gives
\begin{equation}
d_C\alpha = \sum_{s=1}^p F^{\alpha_1}\wedge\dots\wedge F^{\alpha_s}
d v_{\alpha_1\dots \alpha_s} - F^\beta i_\beta \Omega -
\sum_{s=1}^p F^\beta \wedge F^{\alpha_1}\wedge\dots\wedge F^{\alpha_s}
i_\beta v_{\alpha_1\dots \alpha_s}
=0\quad,
\label{3.7}
\end{equation}
and equating equal powers in $F$ the descent equations of Hull and Spence
\cite{Hul.Spe:91} are recovered \cite{Wu:93,Fig.Sta:94}
\begin{equation}
d v_{\alpha_1} = i_{\alpha_1}\Omega\quad,\quad
d v_{\alpha_1\alpha_2}= i_{\{\alpha_2} v_{\alpha_1\}}
\quad,\dots,\quad
d v_{\alpha_1\dots \alpha_s} = i_{\{\alpha_s} v_{\alpha_1\dots \alpha_{s-1}\}}
\quad,
\label{3.8}
\end{equation}
where the symmetrisation,
represented by the curly brackets $\{\ \}$,
is imposed by the commuting $F$'s and includes a factor $1 / s!$.
These equations contain the possible obstructions to the
\emph{problem of gauging} the
form $\Omega$, \ie, to finding an equivariant extension $\widetilde\alpha$
such that
$\mathbf{d}\widetilde\alpha=0,\ \widetilde\alpha \big|_{A=0=F}=\Omega$
(or an $\alpha$ such that $d_C\alpha =0,\ \alpha \big|_{F=0}=\Omega$).

\section{Gauging cocycles on simple groups: general solution}
\label{sec.4}

The descent equations (\ref{3.7}) from the $n$--form $\Omega$
correspond to the pattern
\begin{equation}
\begin{CD}
\Omega^{(0,n)} @>>{i_{\alpha_1}}> (i_{\alpha_1}\Omega^{(0,n)})^{(1,n-1)}
\\
& \mathop{\searrow}\limits_1 & @AA{d}A
\\
& & (v_{\alpha_1})^{(1,n-2)} @>>{i_{\alpha_2}}>
(i_{\{\alpha_2} v_{\alpha_1\}})^{(2,n-3)}
\\
& & & \mathop{\searrow}\limits_2 & @AA{d}A
\\
& & & & (v_{\alpha_1\alpha_2})^{(2,n-4)} & \dots
\\
& & & & & \ddots
\\[0.4cm]
\end{CD}
\end{equation}
Each step $v_{\alpha_1\dots \alpha_s} F^{\alpha_1}\wedge\dots\wedge
F^{\alpha_s} \mapsto v_{\alpha_1\dots \alpha_{s+1}}
F^{\alpha_1}\wedge\dots\wedge F^{\alpha_{s+1}}$
takes a
$(s,n-2s)$--type $n$--form in $\mathcal{W}(\mathcal{H})\otimes\wedge(P)$ to a
$(s+1,n-2s-2)$ one.
We may distinguish two cases:
\begin{description}
\item[a)]
$n$ \emph{odd}, $n$=$2m-1$.
Then, $(m-1)$--steps will lead $\Omega^{(0,2m-1)}$ to $v^{(m-1,1)}$ \ie, to
$v_{\alpha_1 \dots \alpha_{m-1} j} d\varphi^j$.
The $i_{\alpha_m}$ contraction in the $m$-th step will then produce
$
i_{\{\alpha_m} v_{\alpha_1\dots \alpha_{m-1}\}}\equiv c_{\alpha_1\dots\alpha_m}
$
which is a symmetric zero--form.
Then, the last term of eq. (\ref{3.7}) is
\begin{equation}
- c_{\alpha_1\dots\alpha_m} F^{\alpha_1} \wedge\dots\wedge F^{\alpha_{m}}
\quad.
\label{4.1}
\end{equation}
Thus, the form $\alpha$ will be a Cartan cocycle if (\ref{3.6a}) holds and
\begin{equation}
d v_{\alpha_1 \dots \alpha_{s}} =
i_{\{\alpha_s} v_{\alpha_1 \dots \alpha_{s-1}\}}
\quad (s=1,\dots, m-1)
\quad,\quad
i_{\{\alpha_m} v_{\alpha_1\dots \alpha_{m-1}\}}
\equiv c_{\alpha_1\dots\alpha_m} =0
\quad.
\label{4.2a}
\end{equation}
\item[b)]
$n$ \emph{even}, $n=2m$.
In this case, a succession of $m$ steps brings $\Omega^{(0,2m)}$ to $v^{(m,0)}$
\ie, to the $2m$--form $v_{\alpha_1\dots \alpha_m}
F^{\alpha_1}\wedge\dots\wedge F^{\alpha_{m}}$.
Since
$i_\alpha v^{(m,0)}=0$ necessarily, $d_C\alpha$ will be zero
if (\ref{3.6a}) is satisfied and
\begin{equation}
d v_{\alpha_1 \dots \alpha_s} =i_{\{\alpha_s} v_{\alpha_1 \dots \alpha_{s-1}\}}
\quad (s=1,\dots, m)
\quad.
\label{4.2b}
\end{equation}
\end{description}

Finding an $\alpha\in [\mathcal{S}(\mathcal{H}^*)\otimes\wedge(P)]^H$ 
such that eqs. (\ref{4.2a}), (\ref{4.2b}) 
are fulfilled is tantamount to saying that $\Omega$ may be gauged.
This means that we can obtain from $\alpha$ a ($\mathbf{d}$--closed,
gauge invariant) form $\widetilde\alpha$ \cite{Hul.Spe:91}
given by $\psi(\alpha)$, \ie\ by
(\ref{3.5}) with the replacements
$\Omega\to\widetilde\Omega$ and $v_{\alpha_1 \dots \alpha_s} \to
\widetilde{v}_{\alpha_1 \dots \alpha_s}$,
\begin{equation}
\widetilde\alpha = \psi(\alpha)=
\widetilde\Omega + \sum_{s=1}^p {1\over (n-2s)!}
{v}_{\alpha_1 \dots \alpha_s j_1 \dots j_{n-2s}}^{(s,n-2s)}
F^{\alpha_1}\wedge\dots\wedge F^{\alpha_{s}} \wedge
D\varphi^{j_1}\wedge\dots\wedge D\varphi^{j_{n-2s}}
\quad,
\label{4.3}
\end{equation}
where $p=(n-1)/2$ ($n$ odd) or $p=n/2$ ($n$ even). 
For reasons which
will be apparent in a moment, we shall be concerned here with
the odd $n=2m-1$ case only.

Let now $P=G$ where $G$ is a simple, simply connected compact Lie group of
algebra \g\
with basis $\{X_i\}$. We may construct on it WZW
terms on spacetimes of suitable dimension by means of Witten's procedure
\cite{Wit:83} and using the forms on $G$ which define
the Lie algebra cocycles \cite{Che.Eil:48} for each simple \g\ of rank $l$;
they are determined by the $l$ $G$--invariant symmetric polynomials $k$ which
may be constructed on \g\
\footnote{
For details and background references on these topics see, \eg,
\cite{Azc.Mac.Mou.Bue:97,Gre.Hal.Van:76,Azc.Izq:95}.}.
The primitive cocycles are given by the closed\footnote{
For an explicit check see \cite[Lemma 3.1]{Azc.Mac.Mou.Bue:97}.}
\emph{odd} $(2m-1)$--forms on $G$
\begin{equation}
\Omega= k_{i_1 \dots i_{m-1} i_m}
d\omega^{i_1}\wedge \dots \wedge d\omega^{i_{m-1}} \wedge \omega^{i_m}
\quad
(i=1,\dots,\text{dim}\,G)
\quad,
\label{4.4a}
\end{equation}
where $\omega = g^{-1} d g = \omega^{i} T_{i}$
is the left--invariant (LI)
canonical form on $G$ ($\omega^{i}(X_{j})=\delta_j^i$ and $T_i\in\g$
is the generator in the representation of $g$)
and $k_{i_1 \dots i_m}$ is one of the $l$
primitive symmetric invariant polynomials.
We may restrict ourselves to primitive cocycles since they generate the
cohomology ring on $G$.
We may also express (\ref{4.4a}) in the form
\begin{equation}
\Omega \propto
C^{j_1}_{i_1 i_2} \dots C^{j_{m-1}}_{i_{2m-3} i_{2m-2}}
k_{j_1\dots j_{m-1} i_{2m-1}} \omega^{i_1} \wedge \dots \wedge
\omega^{i_{2m-1}}
\quad,
\label{4.4b}
\end{equation}
omitting a factor $(-1/2)^{m-1}$ coming from
$d\omega^i = (-1/2) C^i_{jk} \omega^j \wedge \omega^k$.
The form $\Omega$ (after a suitable pull--back) may be used to define a
$(2m-1)$--dimensional WZW term on a manifold $D$ with a
$(2m-2)$--spacetime $M$ as its boundary, $M=\partial D$ (provided certain
topological conditions are met; we shall not discuss these nor the
quantisation conditions for the WZW term coefficient 
\cite{Wit:83,Kri.Ols.Per:86, DHo.Wei:94}).
We shall now prove the following

\begin{proposition}
\label{prop4.1}
Let $\Omega$ be the closed odd form on a simple, compact and simply connected
Lie group $G$ associated with a primitive cocycle in $H^{2m-1}(\g,\mathbb{R})$.
Let $H$ be a non--trivial Lie subgroup of $G$.
Then the symmetry group $H$ may be gauged if the polynomial $k$ defining
$\Omega$ (eq. (\ref{4.4a})) is zero on its Lie algebra $\mathcal{H}$.
\end{proposition}

\begin{proof}
Since this corresponds to the odd case, we have to show that all conditions
(\ref{4.2a}) are verified by virtue of $\Omega$ being a cocycle.
Introduce now the $[2(m-p)+1]$--forms
on $G$ given by\footnote{
The primes are unnecessary at this stage, but will facilitate in
Sec.~\ref{sec.5} the comparison with the results in \cite{Azc.Mac.Per:98}.}
\begin{equation}
\begin{array}{l}
\displaystyle
\Omega'_{(p)\,\alpha_1\dots \alpha_{p-1}}:=
k_{\alpha_1\dots \alpha_{p-1} i_p i_{p+1} \dots i_{m-1} i_m}
d\omega^{i_p} \wedge \dots \wedge d\omega^{i_{m-1}} \wedge\omega^{i_m}
\displaystyle
\\[0.3cm]
\qquad\qquad\qquad\qquad
\displaystyle
(\alpha=1,\dots,\text{dim}\,H
\quad,\quad
i=1,\dots,\text{dim}\,G)\quad;
\end{array}
\label{4.5}
\end{equation}
$\Omega'_{(p)\,\alpha_1\dots \alpha_{p-1}}$ is symmetric by construction and
$\Omega'_{(1)} = \Omega$.
Clearly if we insert
$F^{\alpha_1} \wedge\dots\wedge F^{\alpha_{p-1}}$ in (\ref{4.5}), the result
is a $(2m-1)$--form in
$\mathcal{S}(\mathcal{H}^*)\otimes \wedge(G)$.
Using that
$i_{X_\alpha} d\omega^i = - C_{\alpha l}^i \omega^l$,
where $X_\alpha\in\mathcal{H}$ is now given by a LI vector field on $G$,
we find
\begin{equation}
i_{\{\alpha_p} \Omega'_{(p)\,\alpha_1\dots \alpha_{p-1}\}} =
-(m-p) C_{\{\alpha_p l}^{i_p} k_{\alpha_1 \dots \alpha_{p-1}\} i_p\dots i_m}
\omega^{l} \wedge \omega^{i_m}\wedge d\omega^{i_{p+1}} \wedge \dots\wedge
d\omega^{i_{m-1}} + \Pi'_{(p)\,\alpha_1 \dots \alpha_p}
\quad,
\label{4.6}
\end{equation}
where we have introduced the $2(m-p)$--form
\begin{equation}
\Pi'_{(p)\,\alpha_1 \dots \alpha_p} \equiv
k_{\alpha_1 \dots \alpha_{p} i_{p+1} \dots i_m}
d\omega^{i_{p+1}} \wedge \dots\wedge d\omega^{i_{m}}\quad.
\label{4.7}
\end{equation}
$\Pi'_{(p)}$ is obviously exact,
\begin{equation}
d \Omega'_{(p)\,\alpha_1\dots \alpha_{p-1}} =
\Pi' _{(p-1)\,\alpha_1 \dots \alpha_{p-1}}
\quad;\quad
\Pi'_{(p)}=\Pi'_{(p)\,\alpha_1 \dots \alpha_p}
F^{\alpha_1} \wedge\dots\wedge F^{\alpha_{p}}
\quad.
\label{4.7a1}
\end{equation}
We now use the $G$--invariance of $k_{i_1 \dots i_m}$ to write
\begin{equation}
\begin{array}{ll}
\displaystyle
-C_{\alpha_p l}^j k_{\alpha_1 \dots \alpha_{p-1} j i_{p+1} \dots i_m}
& = \displaystyle
\sum_{s=1}^{p-1}
C_{\alpha_s l}^j
k_{\alpha_1\dots\widehat{\alpha_s}j\dots\alpha_{p-1}\alpha_p i_{p+1} \dots i_m}
\\
& + \displaystyle
\sum_{s=p+1}^{m-1}
C_{i_s l}^j
k_{\alpha_1\dots\alpha_{p-1}\alpha_p i_{p+1} \dots \widehat{i_s} j \dots i_m}
+ C_{i_m l}^j
k_{\alpha_1\dots\alpha_p i_{p+1} \dots i_{m-1} j}
\quad.
\end{array}
\label{4.7a}
\end{equation}
The second term in the $r.h.s.$ does not contribute to (\ref{4.6}) by the
Jacobi identity.
Symmetrising the $\alpha$'s and using that $k$ is symmetric, eq. (\ref{4.7a})
gives
\begin{equation}
\begin{array}{l}
p C^j_{l\{\alpha_p}
k_{\alpha_1\dots\alpha_{p-1}\} j i_{p+1} \dots i_m}
\omega^{l} \wedge \omega^{i_m}\wedge d\omega^{i_{p+1}} \wedge \dots\wedge
d\omega^{i_{m-1}}
=
\\[0.3cm]
\qquad\qquad\qquad
2 k_{\alpha_1\dots\alpha_p i_{p+1} \dots i_{m}}
d\omega^{i_{p+1}} \wedge \dots\wedge d\omega^{i_{m}} =
2\Pi'_{(p)\,\alpha_1\dots\alpha_p}
\quad.
\end{array}
\label{4.8}
\end{equation}
Thus, eq. (\ref{4.6}) becomes
\begin{equation}
i_{\{\alpha_p} \Omega' _{(p)\,\alpha_1 \dots \alpha_{p-1}\}} =
{2m-p \over p} \Pi'_{(p)\,\alpha_1\dots\alpha_p} =
{2m-p \over p} d \Omega' _{(p+1)\,\alpha_1 \dots \alpha_{p}}
\label{4.9}
\end{equation}
(a rapid way of seeing that exactness holds in each step is to notice that
$i_\alpha [\text{Tr}(T_{\alpha_1}\dots T_{\alpha_{p-1}}\allowbreak
d\omega \wedge \dots \wedge d\omega \wedge \omega)]$
is exact on account of the Maurer--Cartan equations).
The proof is now almost complete: the first steps of the descent are
\begin{equation}
\begin{array}{c}
\displaystyle
i_{\alpha_1} \Omega' _{(1)} = {2m-1 \over 1} d \Omega' _{(2)\, \alpha_1}
\equiv
dv_{\alpha_1}
\quad,
\\[0.3cm]
\displaystyle
i_{\{\alpha_2} v_{\alpha_1\}} =
{2m-1 \over 1} i_{\{\alpha_2} \Omega' _{(2)\, \alpha_1\}} =
{2m-1 \over 1} {2m-2 \over 2} d \Omega' _{(3)\, \alpha_1 \alpha_2} \equiv
d v_{\alpha_1 \alpha_2}
\quad,
\end{array}
\end{equation}
\emph{etc}. Hence, the first set of equations in (\ref{4.2a}) is fulfilled
with
\begin{equation}
v_{\alpha_1\dots\alpha_p} :=
\left( \prod_{r=1}^p {2m-r\over r} \right)
\Omega' _{(p+1)\,\alpha_1 \dots \alpha_{p}}
\quad.
\label{4.10}
\end{equation}
Thus, there is only one possible obstruction to the gauging of $H$, which will
be overcome iff
\begin{equation}
i_{\{\alpha_m} \Omega' _{(m)\,\alpha_1 \dots \alpha_{m-1}\}} =
\Pi'_{(m)\,\alpha_1\dots\alpha_m} = k_{\alpha_1\dots\alpha_m} = 0
\label{4.11}
\end{equation}
\ie, if the polynomial $k_{i_1 \dots i_m}$ on \g\ is zero on
$\mathcal{H}$, \emph{q.e.d.}
Clearly, the group $G$ itself may never be gauged since by hypothesis $k$ is
non--zero on the whole \g\
(but see below and Sec.~\ref{sec.6}).
\end{proof}

The above procedure is a constructive one and, under the sole assumption that
$\Omega$ is a primitive cocycle for \g, provides through (\ref{4.10}) and
(\ref{4.5}) the explicit solution for the form $\alpha$ in (\ref{3.5})
which is a $d_C$--cocycle when Proposition.~\ref{prop4.1} holds.
To find a closed expression for it, let us write
\begin{equation}
\alpha=\sum_{p=1}^{m} \alpha_m(p) \Omega'_{(p)}
\quad,\quad
\alpha_m(p) = \prod_{r=1}^{p-1} {2m-r\over r} \quad, \quad
\Omega'_{(p)} =
\Omega' _{(p)\,\alpha_1 \dots \alpha_{p-1}}
F^{\alpha_1}\wedge\dots\wedge F^{\alpha_{p-1}}\quad.
\label{4.12}
\end{equation}
Hence, and since $d\omega^i = - (\omega\wedge \omega)^i \equiv -(\omega^2)^i$,
the gauged form is given by
$\displaystyle \widetilde\alpha =
\sum_{p=1}^{m} \alpha_m(p) \widetilde \Omega'_{(p)}$
where
\begin{equation}
\begin{array}{l}
\displaystyle
\widetilde \Omega'_{(p)} =
{(-1)^{m-p} \over 2^{m-p}}
k_{\alpha_1\dots \alpha_{p-1} i_p \dots i_{m-1} i_m}
C^{i_p}_{j_{2p-1} j_{2p}} \dots C^{i_{m-1}}_{j_{2m-3} j_{2m-2}}
\\[0.3cm]
\qquad\qquad\qquad
\displaystyle
\cdot F^{\alpha_1}\wedge\dots\wedge F^{\alpha_{p-1}} \wedge
\widetilde\omega^{j_{2p-1}}\wedge\dots\wedge\widetilde\omega^{j_{2m-2}}\wedge
\widetilde\omega^{i_m}
\end{array}
\label{4.13}
\end{equation}
and $\tilde\omega =g^{-1} (d-A^{\alpha} L_{\alpha}) g$.
If we look at the coordinates of the symmetric polynomial in
(\ref{4.13}) as the symmetric trace
$\displaystyle {1\over m!} \text{sTr}
(T_{\alpha_1}\dots T_{\alpha_{p-1}} T_{i_{p}}\dots T_{i_{m}})$,
we may rewrite $\widetilde \Omega'_{(p)}$ as
\begin{equation}
\widetilde \Omega'_{(p)} =
\text{sTr} \{ F^{p-1} \widetilde { {d\omega}^{m-p} {\omega} } \}
= (-1)^{m-p}
\text{sTr} \{ F^{p-1} (\widetilde \omega^2)^{m-p} \widetilde {\omega} \}
\quad.
\label{4.14}
\end{equation}
The symmetric trace over the $m$ factors in (\ref{4.14}) may be replaced by
the trace of the sum
$\mathcal{S}$ over all different `words' which can be made from $(p-1)$ $F$'s
and $(m-p)$ $\omega^2$'s by adding a weight ${(p-1)!(m-p)! / (m-1)!}$
\begin{equation}
\widetilde \Omega'_{(p)}
= (-1)^{m-p} {(p-1)!(m-p)!\over (m-1)!}
\text{Tr} \{ \mathcal{S} [ F^{p-1} (\widetilde \omega^2)^{m-p} ]
\widetilde {\omega} \}
\quad;
\label{4.14a}
\end{equation}
expressions (\ref{4.14}) and (\ref{4.14a}) may be compared with
\cite[eqs. (5.3) and (5.4)]{Azc.Mac.Per:98}.
Thus, the gauged form $\widetilde\alpha$ is given by
\begin{equation}
\begin{array}{rl}
\displaystyle
\widetilde\alpha =
&
\displaystyle
\sum_{p=1}^m (-1)^{m-p}
\alpha_m(p) {(p-1)!(m-p)!\over (m-1)!}
\text{Tr} \{ \mathcal{S} [ F^{p-1} (\widetilde \omega^2)^{m-p} ]
\widetilde {\omega} \}
\\
=
&
\displaystyle
\sum_{p=0}^{m-1} (-1)^{p}
\alpha_m(m-p) {p!(m-p-1)!\over (m-1)!}
\text{Tr} \{ \mathcal{S} [ F^{m-p-1} (\widetilde \omega^2)^{p} ]
\widetilde {\omega} \}
\\
=
&
\displaystyle
\sum_{p=0}^{m-1} (-1)^p {(2m-1)\cdots (m+p+1)(p)!\over (m-1)!}
\text{Tr} \{ \mathcal{S} [ F^{m-p-1} (\widetilde \omega^2)^{p} ]
\widetilde {\omega} \}
\\
=
&
\displaystyle
{(2m-1)!\over (m-1)!m!}
\sum_{p=0}^{m-1} (-1)^p {p!\,m!\over (m+p)!}
\text{Tr} \{ \mathcal{S} [ F^{m-p-1} (\widetilde \omega^2)^{p} ]
\widetilde {\omega} \}
\\
=
&
\displaystyle
{(2m-1)!\over (m-1)!m!}
\sum_{p=0}^{m-1}
\int_0^1 dt\, m\, t^{m-1} (t-1)^{p}
\text{Tr} \{ \mathcal{S} [ F^{m-p-1} (\widetilde \omega^2)^{p} ]
\widetilde {\omega} \}
\end{array}
\label{4.15}
\end{equation}
or, ignoring global factors,
\begin{equation}
\widetilde\alpha \propto \int_0^1 dt\,  \text{Tr}
\Bigl(
\widetilde {\omega} \bigl( t F + t (t-1)(\widetilde \omega^2)\bigr)^{m-1}
\Bigr)
\quad.
\label{4.15a}
\end{equation}
Eq. (\ref{4.15}) provides explicitly the general form of the Weil cocycle
$\widetilde\alpha$. It is formally equivalent to the expression
\cite[eq. (5.8)]{Azc.Mac.Per:98}
of the relative Lie algebra cohomology cocycle
$\bar\Omega^{(2m-1)}\in H^{2m-1}(\mathcal{G},\mathcal{H};\mathbb{R})$,
which may be given as a form on the coset $K=G/H$.
This is now not surprising: it realises the isomorphism
$H^*(\mathcal{G},\mathcal{H};\mathbb{R})=H^*_{\text{DR}}(G/H)=H^*_H(G)$.

If (\ref{4.11}) is not satisfied, $\widetilde\alpha$ will not be closed;
instead (eq. (\ref{4.1})),
\begin{equation}
\mathbf{d}\widetilde\alpha =
-i_{\{\alpha_m} v_{\alpha_2\dots \alpha_m\}}
F^{\alpha_1} \wedge \dots \wedge F^{\alpha_m}
\equiv
-\alpha_m(m) k_{\alpha_1\dots \alpha_m}
F^{\alpha_1} \wedge \dots \wedge F^{\alpha_m}
\quad.
\label{4.16}
\end{equation}
Let us denote by $Q(A,F)$ the Chern--Simons $(2m-1)$--form which is the
local potential of (\ref{4.16}) on $K$,
$dQ(A,F)\propto \text{Tr}(F\wedge\mathop{\cdots}\limits^m\wedge F)\propto
ch_m(F)$ (Chern character).
$Q(A,F)$ has \emph{formally} the same structure as $\widetilde\alpha$
in (\ref{4.15}); in fact, eq. (\ref{4.15}) provides the expression of $Q(A,F)$
if we replace $\widetilde\omega$ by the connection $A$.
Then, the form \cite{Hul.Spe:91}
$\widetilde\alpha'=\widetilde\alpha - Q(A,F)$ will be closed and
hence acceptable for an action leading to $(2m-2)$--dimensional equations of
motion.
However, $\widetilde\alpha'$ will no longer be gauge--invariant due to
$Q(A,F)$;
in fact, $\delta_{\zeta} \widetilde\alpha'$
is proportional to the non--abelian anomaly which is tied to the existence of
the polynomial $k_{i_1\dots i_m}$ which is non--zero on $\mathcal{H}$.

\section{Gauging of forms and effective actions}
\label{sec.5}

Eq. (\ref{4.15}) has the same structure as the general expression
\cite{Azc.Mac.Per:98} which gives
the WZW--type effective actions \emph{\`a la} D'Hoker and Weinberg
\cite{DHo.Wei:94,DHo:95} on the coset $K=G/H$ which
for $G$ simple are obtained from certain cocycles
$\Omega\in\wedge_{2m-1}(G)$.
The key notion in all these constructions is the projectability of forms
(Secs.~\ref{sec.1a},\ref{sec.2}) \ie, their horizontality 
and their $H$--invariance.
Both in the case of gauging WZW terms which have a Lie group $G$ as the
target manifold or in the expressions for the effective actions in
\cite{DHo.Wei:94,DHo:95,Azc.Mac.Per:98}, what matters at the end is the
cohomology of $G/H$ \footnote{
In fact, if the action of a group on a manifold is locally free, the
equivariant and relative (coset) cohomology rings coincide.}.
The Chern--Simons--like \emph{appearance} of the terms of all these formulae
is due to eq. (\ref{4.16}); note, however, that the transgression expression
which gives the Chern--Simons form $Q(A,F)$ of $ch_m(F)$,
$dQ(A,F)\propto \text{Tr}(F\wedge\mathop{\cdots}\limits^m\wedge F)$, is not
projectable.

The WZW effective actions or, equivalently, the relative cohomology cocycles on
the coset $G/H$ are given by \cite[eq. (3.7)]{Azc.Mac.Per:98}
(cf. (\ref{4.12}))
\begin{equation}
\bar\Omega= \sum_{p=1}^m \alpha_m(p) \Omega_{(p)}
\label{5.1aa}
\end{equation}
in terms of the forms \cite[eq. (3.2)]{Azc.Mac.Per:98}
\begin{equation}
\begin{array}{l}
\Omega_{(p)}= (-1)^{p-1} 2^{p-1}
k_{\alpha_1 \dots \alpha_{p-1} i_{p} \dots i_{m-1} b}
C^{i_p}_{a_{2p-1} a_{2p}} \dots C^{i_{m-1}}_{a_{2m-3} a_{2m-2}}
\\[0.3cm]
\qquad\qquad\qquad
\cdot\,\mathcal{W}^{\alpha_1}\wedge\dots \wedge \mathcal{W}^{\alpha_{p-1}}
\wedge \omega^{a_{2p-1}}\wedge\dots \wedge \omega^{a_{2m-2}} \wedge\omega^b
\quad,
\end{array}
\label{5.1a}
\end{equation}
where $\mathcal{W}^{\alpha}=-{1\over 2} C_{ab}^\alpha \omega^a\wedge\omega^b$
is the curvature of the LI $\mathcal{H}$--connection on $G(H,G/H)$
$\omega^\alpha=\mathcal{V}^\alpha$
($\mathcal{W}^{\alpha}=(d\mathcal{V} + \mathcal{V}\wedge\mathcal{V})^\alpha$)
and $a,\ b$ are coset indices in $\mathcal{G}\,\backslash \, \mathcal{H}$.
They are analogous to the present $\Omega'_{(p)}$'s here, eq.
(\ref{4.5}).
The $\Omega_{(p)}$'s in (\ref{5.1a}) satisfy the (Weil model) condition
\cite[eq. (3.5)]{Azc.Mac.Per:98}
\begin{equation}
\mathbf{d}\Omega_{(p)}= -{1\over 2} \Pi_{(p-1)} + {2m-p\over 2p} \Pi_{(p)}
\quad,
\label{5.1}
\end{equation}
where (\cite[eq. (3.6)]{Azc.Mac.Per:98}; cf. (\ref{4.7}))
\begin{equation}
\Pi_{(p)} = (-1) ^{p} 2^{p}
k_{\alpha_1 \dots \alpha_{p} i_{p+1} \dots i_{m}}
C^{i_{p+1}}_{a_{2p+1} a_{2p+2}} \dots C^{i_{m}}_{a_{2m-1} a_{2m}}
\mathcal{W}^{\alpha_1}\wedge\dots \wedge \mathcal{W}^{\alpha_{p}}
\wedge \omega^{a_{2p+1}}\wedge\dots \wedge \omega^{a_{2m}}
\ .
\label{5.1b}
\end{equation}
Notice that, since the exterior derivative in (\ref{5.1}) acts on all forms on
$G$ and hence on $\mathcal{W}^{\alpha}$'s \emph{and} $\omega^{a}$'s in
(\ref{5.1a}), it corresponds to $\mathbf{d}$ in the notation of this paper.

We see that we can relate $\Omega_{(p)},\ \Pi_{(p)}$
(relevant in the analysis of the effective actions in
\cite{Azc.Mac.Per:98}) with their $\Omega'_{(p)},\ \Pi'_{(p)}$ 
counterparts (used here in the analysis of the gauged
WZW terms) by means of the replacements
$\mathcal{W}^\alpha\to F^\alpha,\ \omega^a \to \widetilde\omega^i$.
Hence, we move from the expressions of the effective actions in
\cite{Azc.Mac.Per:98} to those of the present WZW
gauged terms by the replacements
$\mathcal{W}^\alpha\mapsto F^\alpha,\
{1\over 2} C^i_{ab}\omega^a\wedge\omega^b\mapsto
{1\over 2} C^i_{jk}\widetilde\omega^j\wedge\widetilde\omega^k$
\footnote{The $\mathcal{H}$--horizontal  forms $\omega^a$
($i_{X_\alpha} \omega^a=0$, $\alpha$ in $\mathcal{H}$, $a$ in the coset
$\mathcal{G}\,\backslash\,\mathcal{H}$), 
appearing in the relative cohomology cocycles 
which define effective actions, find their counterparts
here in $\psi(\omega^i)=\widetilde\omega^i=[g^{-1}(d-A^\beta L_\beta)g]^i$.
The $\widetilde\omega^i$ (in contrast with $\omega^i$) are horizontal by
construction; explicitly,
$\mathbf{i}_\alpha ( g^{-1} (d - A^\beta L_\beta) g ) =
( g^{-1} (L_\alpha - d i_\alpha - L_\alpha) g ) = 0$.}.
The reason for this common structure may be also understood 
in terms of the equivariant cohomology (in the Cartan model, 
l.h.s. of diagram (\ref{3.4b})).
In the problem of gauging WZW forms discussed in the previous
sections, the Weil algebra
$\mathcal{W}(\mathcal{H})$ is generated by the gauge fields
$A^\alpha$ and the curvature $F^\alpha$. Therefore, the forms $\omega$
spanning the exterior algebra $\wedge(G)$ are (minimally) coupled by
$\psi (\omega)=\tilde\omega$ to $A$.
However, the structure of the Cartan model expressions
does not depend on the specific connection and curvature, 
and the above effective actions also correspond to equivariant
cocycles. The only difference is that, when 
we are interested in effective actions and relative
cohomology, we take as generators of the Weil algebra
$\mathcal{W}(\mathcal{H})$ the $\mathcal{H}$--connection
$\mathcal{V}^\alpha$ and its curvature $\mathcal{W}^\alpha$.
The `minimal coupling' analogue to (\ref{3.2}) 
is then given by
\begin{equation}
\psi (\omega) = \prod_\alpha (1-\mathcal{V}^\alpha i_\alpha)\omega =
\omega - \mathcal{V} \equiv \mathcal{U}
\quad,
\label{5.5a}
\end{equation}
where $\mathcal{U}$ is, clearly, the coset or
$(\mathcal{G}\,\backslash\,\mathcal{H})$--component of the LI canonical form
$\omega = \omega\big|_\mathcal{H} + \omega\big|_\mathcal{K} \equiv
\mathcal{V} + \mathcal{U}$.

This explains the similarity of the final expressions. 
For instance, eq. (\ref{5.1}) may be computed in the Cartan model.
Using that $\psi^{-1}(\Omega_{(p)})=\Omega_{(p)}(\mathcal{V}=0)$ we see that
\begin{equation}
\psi^{-1}(\Omega_{(p)})=(-1)^{m-1} 2^{m-1}
\Omega'_{(p)\,\alpha_1\dots\alpha_{p-1}}\mathcal{W}^{\alpha_1}
\dots\mathcal{W}^{\alpha_{p-1}}
\quad.
\end{equation}
Thus, with
(\ref{4.5}) and (\ref{4.9}) we find
\begin{equation}
\begin{array}{l}
\displaystyle
 d_C \left( \psi^{-1}(\Omega_{(p)}) \right)=
d_C \left ( (-1)^{m-1} 2^{m-1} \Omega'_{(p)\,\alpha_1\dots \alpha_{p-1}}
\mathcal{W} ^{\alpha_1}\wedge\dots\wedge \mathcal{W} ^{\alpha_{p-1}} \right)=
\\[0.3cm]
\displaystyle
=
(-1)^{m-1} 2^{m-1} \Bigl( \Pi'_{(p-1)\,\alpha_1\dots \alpha_{p-1}}
\mathcal{W} ^{\alpha_1}\wedge\dots\wedge \mathcal{W} ^{\alpha_{p-1}} -
{2m-p\over p} \Pi'_{(p)\,\alpha_1\dots \alpha_{p}}
\mathcal{W} ^{\alpha_1}\wedge\dots\wedge \mathcal{W} ^{\alpha_{p}}
\Bigr)
\end{array}
\label{5.3}
\end{equation}
which corresponds to (\ref{5.1}) once $\psi$ has been applied to it
(note that, for (\ref{5.5a}),
$\psi \Pi'_{(p)} = {1\over (-1)^m 2^m}\Pi_{(p)}$).
This shows that the Cartan derivative $d_C$ is equivalent to $\mathbf{d}$
once the minimal coupling has been performed.

Summarising, we may express these results in the following general form
\begin{theorem}
Let $\Omega$ be a closed $(2m-1)$--Lie algebra cocycle given by a
$(2m-1)$--form on the manifold $G$, and
let $H$ be the structure group of the principal bundle $G(H,K)$, $K=G/H$.
Let $F\ (\text{resp.}\ \mathcal{W})$ be the curvature associated with $A$
(resp. with the LI $\mathcal{H}$--connection $\mathcal{V}$).
Then it will be possible to construct from $\Omega$ a) an effective action
$\bar\Omega$ on the coset manifold $K$
and b) an $H$--gauged, closed and gauge invariant form $\widetilde\alpha$
if $\Omega$ is a Lie algebra cocycle in $H^{2m-1}(\g,\mathbb{R})$ defined by a
symmetric invariant polynomial which vanishes on $\mathcal{H}$.
In this case, $\bar\Omega$ (eq. (\ref{5.1aa})) and $\alpha$ (eq. (\ref{4.12}))
will be respectively, cocycles in the relative Lie algebra
$H^{2m-1}(\g,\mathcal{H};\mathbb{R})$ and
equivariant $H^{2m-1}_H(G)$ cohomologies.
\end{theorem}

The above constructions constitute, in fact, a physics inspired
proof of the isomorphism between the relative and equivariant
(for the action of $H$ on $G$) cohomologies.
The obstruction to constructing the effective action and to gauging the WZW
term has the same geometrical origin; it is given in terms of 
an anomaly, which appears
when $k_{\alpha_1\dots\alpha_m}$ is non--zero.

\section{The gauging of left and right symmetries}
\label{sec.6}
It was stated in Sec.~\ref{sec.4} that $G$ itself may never be gauged.
But being $G$
a compact group the cocycles $\Omega$ on $G$ are \emph{both} LI and
right invariant (RI): there is a $G^L\times G^R$ symmetry.
Thus, although a simple factor $G$ may not be gauged, we may expect to have the
unwanted contributions to $\mathbf{d}\widetilde{\alpha}$ from each factor to
cancel each other.
Moreover, even if $k_{\alpha_1\dots\alpha_m}$ is non-zero on $\mathcal{H}$,
we may use this fact to overcome the obstruction which would be
present for $H\times 1$ or $1\times H$ separately.
We shall now do this and provide general expressions for the gauged WZW terms
following the above pattern.

The $R$ and $L$ actions are generated, respectively, by the LI and RI vector
fields $X_{\alpha}^{L}$ and $-X_{\alpha}^{R}$ (the sign is introduced to
compensate for the -- sign in
$[X_{\alpha}^{R},X_{\beta}^{R}]= -C_{\alpha\beta}^\gamma X_{\gamma}^{R}$ which
is required if we conventionally adopt $[X^L_\alpha,X^L_\beta]=
C_{\alpha\beta}^\gamma X^L_\gamma$).
Let us denote by $i_{L\,\alpha}$ the inner product $i_{X_{\alpha}^{L}}$
and by $i_{R\,\alpha}$ the inner product $i_{(-X_{\alpha}^{R})}$.
Let $(A_L^\alpha,F_L^\alpha)$ and $(A_R^\alpha,F_R^\alpha)$ be two different
copies of the Weil algebra $\mathcal{W}(\mathcal{H})$ (the indices L and R in
$(A,F)$ correspond to the accompanying LI or RI vector fields and hence to the
R
and L actions).
Then,
\begin{equation}
F^\alpha_L i_{L\,\alpha} \omega = F_L
\quad,\quad
F^\alpha_R i_{R\,\alpha} \omega = - {g^{-1} F_R g}
\quad.
\label{6.1}
\end{equation}
In the sequel, the following relations will be useful
\begin{equation}
F^\alpha_L i_{L\,\alpha} \omega^2 = [F,\omega] \quad,\quad
F^\alpha_R i_{R\,\alpha} \omega^2 = - [ {g^{-1} F_R g}, \omega]\quad,
\label{6.2}
\end{equation}
\begin{equation}
(1\otimes d) ({g^{-1} F_R g}) \equiv d ({g^{-1} F_R g}) = [{g^{-1} F_R
g},\omega] \quad.
\label{6.3}
\end{equation}
Let us now introduce the $(2m-1)$--forms ${\Upsilon}_{{[p,q]}}$
and the $(2m)$--forms ${\Pi}_{{[p,q]}}$ and ${\Gamma}_{{[p,q+1]}}$
by\footnote{
\label{foot6}
The previous forms $\Omega'_{(p)}$, $\Pi'_{(p)}$ are particular cases of
${\Upsilon}_{{[p,q]}}$, ${\Pi}_{{[p,q]}}$:
${\Upsilon}_{{[0,0]}}=(-1)^{m-1}\Omega$ (eq. (\ref{4.4a})) and, for
$F_L\equiv F$, ${\Upsilon}_{{[p,0]}} = (-1)^{m-p-1} \Omega'_{(p+1)}$
(cf. (\ref{4.14})) and ${\Pi}_{{[p,0]}}= (-1)^{m-p} \Pi'_{(p)}$ (cf.
(\ref{4.7})).}
\begin{equation}
\begin{array}{l}
\displaystyle
{\Upsilon}_{{[p,q]}} = {\text{sTr}}(\omega F_L^p ({g^{-1} F_R g})^q
(\omega^2)^{m-p-q-1}) \quad,
\\[0.25cm]
\displaystyle
{\Pi}_{{[p,q]}} = {\text{sTr}} (F_L^p ({g^{-1} F_R g})^q
(\omega^2)^{m-p-q})\quad,
\\[0.25cm]
\displaystyle
{\Gamma}_{{[p,q+1]}} = {\text{sTr}} (\omega F_L^p ({g^{-1} F_R g})^q
[{g^{-1} F_R g},\omega] (\omega^2)^{m-p-q-2})\quad.
\end{array}
\label{6.4}
\end{equation}
We show in the Appendix that
\begin{equation}
{\Gamma}_{[0,q+1]} = - {2\over q+1} {\Pi}_{[0,q+1]} \quad,
\label{6.5}
\end{equation}
\begin{equation}
{\text{sTr}} (\omega F_L^p [F_L,\omega] ({g^{-1} F_R g})^q
(\omega^2)^{m-p-q-2})
= - {1\over p+1} (2 {\Pi}_{[p+1,q]} + q {\Gamma}_{[p+1,q]})
\quad.
\label{6.6}
\end{equation}

To find now the equivariant extension of $\Omega$ we need to compute
$d_C {\Upsilon}_{{[p,q]}}$, and hence the action of $d$,
$F_L^\alpha i_{L\,\alpha}$ and $F_R^\alpha i_{R\,\alpha}$ on
${\Upsilon}_{{[p,q]}}$.
It is not difficult to check, using (\ref{6.1}), (\ref{6.2}), (\ref{6.3}) and
(\ref{6.6}), that
\begin{equation}
d {\Upsilon}_{{[p,q]}} = - {\Pi}_{{[p,q]}} -q {\Gamma}_{[p,q]}
\quad,
\label{6.7a}
\end{equation}
\begin{equation}
F_L^\alpha i_{L\,\alpha} {\Upsilon}_{{[p,q]}} = {\Pi}_{[p+1,q]} +
{m-p-q-1\over p+1} ( 2 {\Pi}_{[p+1,q]} + q {\Gamma}_{[p+1,q]} )
\quad,
\label{6.7b}
\end{equation}
\begin{equation}
F_R^\alpha i_{R\,\alpha} {\Upsilon}_{{[p,q]}} = - {\Pi}_{[p,q+1]} +
(m-p-q-1) {\Gamma}_{{[p,q+1]}}
\quad.
\label{6.7c}
\end{equation}
Now, let us introduce ($s\ge1$)
\begin{equation}
{\Omega}_{[s]} := \sum_{p+q=s-1} {(m-p-1)! (m-q-1)! \over p! q!}
{\Upsilon}_{{[p,q]}}
\quad,\quad \Omega_{[1]} = (-1)^{m-1} (m-1)!^2 \Omega
\quad.
\label{6.8}
\end{equation}
Using (\ref{6.7a}), (\ref{6.7b}) and (\ref{6.7c}) we find (see (\ref{A.3}))
\begin{equation}
(F_L^\alpha i_{L\,\alpha} + F_R^\alpha i_{R\,\alpha} ) {\Omega}_{[s]}
=
-(2m-s)(m-s) d {\Omega}_{[s+1]}
\quad.
\label{6.9}
\end{equation}

We now observe that eq. (\ref{6.9}) has the same structure as (\ref{4.9}).
Hence, the cohomological descent shows that the form
\begin{equation}
\alpha = \sum_{s=1}^m { (-1)^{m-s} \over (2m-s)!(m-s)! }
\Omega_{[s]}
\label{6.10}
\end{equation}
verifies ($d\Omega_{[1]} =0$)
\begin{equation}
\begin{array}{rl}
d_C\alpha
\equiv & \displaystyle
(d - F_L^\alpha i_{L\,\alpha} - F_R^\alpha i_{R\,\alpha} ) \alpha
= \sum_{s=2}^{m} { (-1)^{m-s} \over (2m-s)!(m-s)! }
d \Omega_{[s]}
\\
+ & \displaystyle
\sum_{s=1}^{m-1}
{ (-1)^{m-s} \over (2m-s-1)!(m-s-1)! } {\Omega}_{[s+1]}
- ( F_L^\alpha i_{L\,\alpha} + F_R^\alpha i_{R\,\alpha} )
{1\over m!} {\Omega}_{[m]}
\\
= & \displaystyle
- ( F_L^\alpha i_{L\,\alpha} + F_R^\alpha i_{R\,\alpha} )
{1\over m!} {\Omega}_{[m]}
\quad.
\end{array}
\label{6.11}
\end{equation}
Now, using (\ref{A.4})
\begin{equation}
( F_L^\alpha i_{L\,\alpha} + F_R^\alpha i_{R\,\alpha} ) {\Omega}_{[m]}
= {\Pi}_{[m,0]} - {\Pi}_{[0,m]}
\equiv
{\text{sTr}} ( F_L^m) - {\text{sTr}} ( ({g^{-1} F_R g})^m )
\quad,
\label{6.12}
\end{equation}
so that
\begin{equation}
d_C\alpha =
- {1\over m!} \Big( {\text{sTr}} ( F_L^m) - {\text{sTr}} ( ({g^{-1} F_R g})^m )
\Big)
=
- {1\over m!} \Big( {\text{sTr}} (F_L^m) - {\text{sTr}} (F_R^m) \Big)
\label{6.13}
\quad.
\end{equation}
In particular, if $F_L = F_R \equiv F$ eq. (\ref{6.13}) is zero and $\alpha$
is an equivariant cocycle in the Cartan model.
We may then state the following

\begin{proposition}
Let $\Omega$ be a cocycle on $G$, $H\times H$ the symmetry to be gauged.
Then, the extension $\alpha$ of $\Omega$ given by
\begin{equation}
\alpha = \sum_{s=1}^m \sum_{p+q=s-1}
{(-1)^{m-s} (m-p-1)! (m-q-1)! \over (2m-s)!(m-s)! p! q! }
{\text{sTr}}(\omega F_L^p ({g^{-1} F_R g})^q (\omega^2)^{m-p-q-1})
\label{6.14}
\end{equation}
is an equivariant extension of $\Omega$ for $F_L=F_R=F$.
\end{proposition}

To make contact with the work in \cite{Hul.Spe:91} let us note that
the symmetric trace and the double sum may be replaced by a `trinomial'
using that
\begin{equation}
\text{Tr} ( F_L + {g^{-1} F_R g} + (\omega^2) )^{m-1} =
\sum_{p,q}  {1\over p! q! (m-p-q-1)! }
\text{sTr} ( F_L^p ({g^{-1} F_R g})^q (\omega^2)^{m-p-q-1})
\quad.
\label{6.15}
\end{equation}
Thus, splitting the sum over $p$ and $q$ in a sum over $s\equiv p+q+1$ and
a sum over $s$ and recalling that the Beta function
\begin{equation}
B(v+1,w+1)=
\int_0^1 dt \, t^v (1-t)^w = {v! w!\over (v+w+1)!}
\quad,
\label{6.16}
\end{equation}
we obtain the expression
\begin{equation}
\begin{array}{c}
\displaystyle
\sum_{s=1}^m \sum_{p+q=s-1}
{(-1)^{m-s} \over (2m-s)!(m-s)!} { (m-p-1)! (m-q-1)! \over p! q!}
{\text{sTr}}(\omega F_L^p ({g^{-1} F_R g})^q (\omega^2)^{m-p-q-1})
\\
\displaystyle
=
\int_0^1 dt \, \text{Tr} \Big( \omega \big( t F_L + (1-t) {g^{-1} F_R g} +
t(t-1) \omega^2 \big)^{m-1} \Big)
\quad,
\end{array}
\label{6.17}
\end{equation}
which recovers
\cite[eq. (7.41) (ignoring the Chern--Simons terms there)]{Hul.Spe:91}
once the minimal coupling has been performed.
For $1\times H\equiv H$, $F=F_L$, $F_R=0$, we obtain eq. (\ref{4.15a})
(before minimal coupling).

In the present framework, the minimal coupling is implemented by means of the
gauging map (\ref{3.2}) which here (for $A_L \ne A_R$) takes the form
\begin{equation}
\psi = \prod_\alpha (1 - A_L^\alpha i_{L\,\alpha} - A_R^\alpha i_{R\,\alpha})
\quad,
\label{6.18}
\end{equation}
so that
\begin{equation}
\widetilde{\omega} = \psi(\omega) = \omega - A_L + g^{-1} A_R g
=g^{-1} ( d - g A_L g^{-1}  + A_R ) g \equiv
g^{-1} D g
\label{6.19}
\quad.
\end{equation}
For $A_L=A$, $A_R=0$, eq. (\ref{6.19}) reduces to $D=d-A^\alpha L_{\alpha}$
(eq. (\ref{2.1})).

\begin{example}
Let us illustrate the above with the lowest example for $A_L=A_R=A$,
$F_L=F_R=F$ (the three-cocycle, $m=2$)
\cite{Hul.Spe:91,Wit:92b}.
In this case, our expression above
has three terms, one corresponding to $s=1$ ($p=q=0$) and two for
$s=2$ ($p=1,\ q=0$ and $p=0,\ q=1)$. Explicitly
(cf. \cite[eq. (7.18)]{Hul.Spe:91}),
\begin{equation}
\alpha= -{1\over 3!} {\text{sTr}}(\omega (\omega^2) )
+{1\over 2!} {\text{sTr}}(\omega F) +
{1\over 2!} {\text{sTr}}(\omega g^{-1} F g )
\quad,
\end{equation}
where the first term corresponds to the original three-cocycle
$\Omega\propto \text{Tr}(\omega^3)$.
Substituting $\omega$ by its gauged version
$\widetilde\omega=\omega -A + g^{-1} A g$ (cf. (\ref{6.19})) we obtain
the gauged WZW action in a two--dimensional spacetime.
\end{example}

\section{Concluding remarks}

WZW terms on a simple group $G$ as the target manifold are obtained from odd
forms $\Omega\in H^{2m-1}(\g,\mathbb{R}),\ m\ge 2$, which define
non--trivial primitive cocycles in the Lie algebra cohomology.
These are in turn characterised by primitive symmetric invariant polynomials
of order $m$.
In this paper we have given a closed and general expression  for the forms
$\alpha$ which provide the $H$--gauged version or `extension' $(H\subset G)$
of such Lie algebra cocycles $\Omega$.
This expression is explicitly constructed from the invariant polynomials
on \g\ (which are all known for $\g$ simple).
We have also explained the similarity between the gauged extensions of the
various $\Omega$'s and the expression of the
WZW--type $G$--invariant effective actions of D'Hoker and Weinberg 
relative to the coset $G/H$. The correspondence 
among these two types of action terms constitutes a physical 
realisation of the mathematical isomorphism 
$H_H^*(G)\sim H^*(\g,\mathcal{H};\mathbb{R})$ between the 
equivariant and relative cohomologies.

Since we have been concerned here with simple algebras, only odd forms $\Omega$
on $G$ have entered into our discussion since the \emph{primitive}
$2m$--cocycles on $\g$ are coboundaries (exact forms on $G$).
We might, of course, remove the semisimplicity condition.
The cohomology theory in the non--semisimple case, however, is not complete,
so that a general constructive process (similar to the one presented here) does
not exist (nevertheless, we wish to mention here that the
contraction of Lie algebras may provide a systematic procedure to
discuss the cohomology of non--semisimple algebras, a first step 
to extend the physical considerations of the present paper).
Also, in the non--semisimple case it is possible to introduce non--trivial
cocycles which take values in a representation space $V$ of \g, 
\ie, elements in $H_\rho^k(\g,V)$ where $\rho$ is a representation of
\g\ (by the Whitehead Lemma, 
$H_\rho^k(\g,V)=0\ \forall\,k\ge 0$ if $\rho$ is non--trivial and
\g\ is semisimple).
In particular, this approach might lead to 
a different class of topological terms \cite{deW.Hul.Roc:87} 
(which are not obtained by gauging a WZW term $\Omega$ and hence 
are zero for $A=0=F$) in even dimensional spacetime and which may be
added to the kinetic term for gauge theories with noncompact groups.
We may come back to these problems and to their extension to the
supersymmetric case (see \eg, \cite{Hul.Kar.Lin.Roc:86,Sho:89})
in the future.

\begin{noteadded}
On the subject of the topological terms in \cite{deW.Hul.Roc:87} we have just
become aware of \cite{Hen.Wil:98}.
\end{noteadded}

\subsection*{Acknowledgements}
This paper has been partially supported by a research grant from
the MEC, Spain (PB96-0756).
J.~C.~P.~B. wishes to thank the Spanish MEC and the CSIC for an FPI grant.
A conversation with G. Papadopoulos and helpful comments of J. Stasheff on the
manuscript are gratefully acknowledged.

\section*{Appendix}
\renewcommand{\theequation}{A.\arabic{equation}}
\setcounter{equation}{0}

We prove here some expressions used in Sec.~\ref{sec.6}.
Eq. (\ref{6.5}) may be derived using that the forms
$\Upsilon_{[p,q]}$ are defined by a symmetrised trace.
Thus,
\begin{equation}
\begin{array}{r@{}l}
0 =
\text{sTr}([\omega (g^{-1} F_R g)^{q+1} &
                                          (\omega^2)^{m-q-2},\omega])
=
\text{sTr}([\omega,\omega] (g^{-1} F_R g)^{q+1} (\omega^2)^{m-q-2})
\\[0.25cm]
+ &
(q+1)\text{sTr}
(\omega (g^{-1} F_R g)^{q} [g^{-1} F_R g,\omega](\omega^2)^{m-q-2})
\\[0.25cm]
+ &
(m-q-2)\text{sTr}
(\omega (g^{-1} F_R g)^{q+1} (\omega^2)^{m-q-3} [\omega^2,\omega])
\end{array}
\end{equation}
which, since $[\omega^2,\omega]=0$ (Jacobi), reduces to eq. (\ref{6.5})
\begin{equation}
2\Pi_{[0,q+1]} + (q+1) \Gamma_{[0,q+1]} =0
\quad.
\label{A.2}
\end{equation}
Eq. (\ref{6.6}) follows similarly from
$\text{sTr}([\omega F_L^{p+1} (g^{-1} F_R g)^{q} (\omega^2)^{m-p-q-2},\omega])
=0$.

The relations (\ref{4.9}) (used to find $\alpha$
when gauging the (left) symmetry group $H$) may be recovered from
(\ref{6.7a}) and (\ref{6.7b}).
For $q=0$, these equations give
\begin{equation}
d {\Upsilon}_{{[p,0]}} = - {\Pi}_{{[p,0]}}
\quad,\quad
F_L^\alpha i_{L\,\alpha} {\Upsilon}_{{[p,0]}} = {\Pi}_{[p+1,0]} +
{2(m-p-1)\over p+1} {\Pi}_{[p+1,0]}
\quad.
\label{A.2a}
\end{equation}
For $F_L\equiv F$, and recalling (see footnote~\ref{foot6}) that
${\Upsilon}_{{[p,0]}}=(-1)^{m-p-1}\Omega'_{(p+1)}$,
${\Pi}_{{[p,0]}}=(-1)^{m-p}{\Pi'}_{{(p)}}$ we find
\begin{equation}
d \Omega'_{(p+1)} = {\Pi'}_{{(p)}}
\quad,\quad
F^\alpha i_{\alpha} \Omega'_{(p+1)} = {2m-p-1 \over p+1} {\Pi'}_{{(p+1)}}
\label{A.2b}
\end{equation}
which reproduce (\ref{4.9}).
An equivalent result is obtained for the right symmetry group by simply setting
$p=0$ in (\ref{6.7a}) and (\ref{6.7c})
\begin{equation}
d {\Upsilon}_{{[0,q]}} = - {\Pi}_{{[0,q]}} - q\Gamma_{{[0,q]}}
\quad,\quad
F_R^\alpha i_{R\,\alpha} {\Upsilon}_{{[0,q]}} = - {\Pi}_{{[0,q+1]}}
+(m-q-1) \Gamma_{{[0,q+1]}}
\quad.
\label{A.2c}
\end{equation}
Then, using (\ref{A.2}) (cf. (\ref{A.2a}))
\begin{equation}
d {\Upsilon}_{{[0,q]}} = {\Pi}_{{[0,q]}}
\quad,\quad
F_R^\alpha i_{R\,\alpha} {\Upsilon}_{{[0,q]}} = - {\Pi}_{{[0,q+1]}}
- {2(m-q-1)\over q+1} {\Pi}_{{[0,q+1]}}
\quad.
\label{A.2d}
\end{equation}

Eq. (\ref{6.9}) may be computed as follows
\begin{equation}
\begin{array}{r@{}l}
\displaystyle
(&F_L^\alpha i_{L\,\alpha} + F_R^\alpha i_{R\,\alpha} ) {\Omega}_{[s]}
=
\displaystyle
\sum_{{p+q=s-1}} {(m-p-1)! (m-q-1)! \over p! q!}
\Big[ \left(1+{2(m-s)\over p+1}\right) \Pi_{[p+1,q]}
\\[0.3cm]
&
\displaystyle
\qquad\qquad\qquad\qquad
+ {q(m-s)\over p+1} \Gamma_{[p+1,q]}
- \Pi_{[p,q+1]} + (m-s) \Gamma_{[p,q+1]} \Big]
\\[0.3cm]
& =
\displaystyle
{(m-s-1)! (m-1)!\over s!} \Big\{
(2m-s)(m-s) {\Pi}_{[s,0]}
-
s (m-s) ({\Pi}_{[0,s]} - (m-s)\Gamma_{[0,s]}) \Big\}
\\[0.3cm]
&
\displaystyle
+
\sum_{\scriptstyle {p+q=s-1}\atop \scriptstyle p\ne 0,q\ne s-1}
{(m-p-1)! (m-q-2)! \over p! (q+1)!}
\Big\{\Big [(m-p)(2m-2s+p)
\\[0.3cm]
& -
\displaystyle
(m-q-1)(q+1)\Big] {\Pi}_{[p,q+1]}
+
(q+1) \Big [ (m-p)(m-s) + (m-q-1) (m-s) \Big] {\Gamma}_{[p,q+1]} \Big\}
\\[0.3cm]
& =
\displaystyle
(2m-s)(m-s)
\Big\{
{(m-s-1)! (m-1)!\over s!} ({\Pi}_{[s,0]} - {\Pi}_{[0,s]})
\\[0.3cm]
&
\displaystyle
\qquad\qquad
\qquad\qquad +
\sum_{\scriptstyle {p+q=s-1}\atop \scriptstyle p\ne 0,q\ne s-1}
{(m-p-1)! (m-q-2)! \over p! (q+1)!}[ {\Pi}_{[p,q+1]} + (q+1) {\Gamma}_{[p,q+1]}
]
\Big\}
\\[0.3cm]
& =
\displaystyle
-(2m-s)(m-s)
d \Big\{
{(m-s-1)! (m-1)!\over s!} ({\Upsilon}_{[s,0]} + {\Upsilon}_{[0,s]})
\\[0.3cm]
&
\displaystyle
\qquad\qquad
\qquad\qquad +
\sum_{\scriptstyle {p+q=s-1}\atop \scriptstyle p\ne 0,q\ne s-1}
{(m-p-1)! (m-q-2)! \over p! (q+1)!} {\Upsilon}_{[p,q+1]}
\Big\}
\\[0.3cm]
& =
-(2m-s)(m-s) d {\Omega}_{[s+1]}
\quad,
\end{array}
\label{A.3}
\end{equation}
where we have used (\ref{6.7b}), (\ref{6.7c}) in the first equality, then we
have changed the summation indices $p+1\to p$ and $q\to q+1$ in the first and
second terms to obtain the second equality, and eq. (\ref{A.2}) has been
used in the third one. Finally, eq. (\ref{6.7a}) and the first
identity in (\ref{A.2d}) lead to the fourth equality which trivially rearranges
into $-(2m-s)(m-s)d {\Omega}_{[s+1]}$.

We also note that the first equality of this calculation also shows that
\begin{equation}
( F_L^\alpha i_{L\,\alpha} + F_R^\alpha i_{R\,\alpha} ) {\Omega}_{[m]}
\sum_{{p+q=s-1}} {(m-p-1)! (m-q-1)! \over p! q!}
\Big[
\Pi_{[p+1,q]} - \Pi_{[p,q+1]}
\Big]
= {\Pi}_{[m,0]} - {\Pi}_{[0,m]}
\ .
\label{A.4}
\end{equation}


\end{document}